# Enabling temperature controlled *in-situ* vapor dosing for lab source X-ray reflectivity measurements

*Erin E. Dunphy[1], Mara A. Fischer[1], Sasha R. Neefe[1], Devin L. Shaffer[2], Anthony P. Straub[3,4], Michael F. Toney[1,4,5*], J. Will Medlin[1*]*

[1]Department of Chemical & Biological Engineering, University of Colorado Boulder, Boulder, Colorado 80303, United States

[2]Department of Civil and Environmental Engineering, University of Houston, Houston, Texas 77204, USA

[3]Department of Mechanical and Process Engineering (D-MAVT), ETH Zurich, 8092 Zurich, Switzerland

[4]Materials Science & Engineering Program, University of Colorado, Boulder, Colorado 80303, United States

[5]Renewable & Sustainable Energy Institute, University of Colorado, Boulder, Colorado 80303, United States

**Abstract**

X-ray Reflectivity (XRR) is a valuable technique for probing buried interfaces in complex systems relevant to thin-film, membrane, and battery applications, among others. However, many *operando* and *in situ* reflectivity cells are designed for use at synchrotron facilities, limiting the broader accessibility of these measurements. We present an XRR transmission cell that enables *in-situ* vapor dosing and temperature-dependent experiments on in-house diffractometers. We demonstrate its capabilities with two case studies: the adsorption of water into polyamide (PA) membranes on silicon and temperature-dependent restructuring of polystyrene (PS) pseudo-brushes on alumina. Vapor dosing allows for controlled release of vapor into the cell, allowing operation across a wide range of conditions from rough vacuum to saturation. We demonstrate that the manifold can reach 90-95% of saturated pressures, with the measurements presented here spanning 0-250 mbar, which is desirable for adsorption isotherms. Heating studies performed between 25 and 200ºC demonstrate the ability to resolve Ångstrom scale structural changes in a surface bound polymer. These results establish a novel streamlined approach to temperature-controlled vapor dosing on a laboratory diffractometer, offering straightforward probe-molecule exchange, vacuum-sealed operations, and variable temperature capabilities.

## I. Introduction

Over the last four decades the use of X-ray reflectivity (XRR) has become a critical method for probing dynamic interfaces. XRR utilizes a specular grazing incident geometry and enables extraction of electron density profiles perpendicular to the surface. The electron density profiles define the thickness, electron-density, and roughness of each distinct layer[1]. Ultimately, XRR is ideal for studying the evolution of complex and buried interface films *in situ* and has been employed to characterize various membrane[2,3], catalytic[2], mineral[3,4], Langmuir and polymer thin film[5–9], and battery[10–12] materials.

Many reflectivity cells have been designed to study liquid-solid and gas-solid interfaces under a range of temperature and pressure conditions. Fenter discusses two standard reflectivity cell types for the characterization of interfaces: a thin-film cell and a transmission cell[4]. A thin-film cell typically operates by trapping liquid or gas between the substrate and a plastic film (often Kapton). Using a thin film minimizes background scattering during the measurement. However, there are several limitations: the composition of the thin liquid film may change during the measurement if the substrate is reactive, gas may permeate through the plastic film window, and kinetic studies are challenging. Thum et al. developed a complex thin-film cell capable of probing a Pd thin film interface during hydrogenation at 150ºC under ambient pressures[2]. For these reaction studies, precise gas control over a small sample area (5 mm x 10 mm) enabled kinetic measurements where gas was swiftly replaced in the 5 mm space above the substrate. Construction of thin-film cells requires precise machining and substrate consistency (e.g. thickness), and there is a possibility of reflectivity from the plastic film surface interfering with that from the interface of interest. Additionally, thin-film cells are typically purpose-built with a specific substrate or chemical systems in mind, which limits their broader applicability.

Transmission cells differ from thin-film cells in that they contain a macroscopically thick solution or vapor phase above the substrate surface. This enables a more controlled sample environment. Transmission cells are typically larger than thin-film cells and fill the volume above the surface completely with liquid or vapor, increasing the relative background for reflectivity scans. However, the advantage of a larger cell is that there is more room to add functionality. Transmission cells can become quite intricate, like the hanging-meniscus cell developed by Magnussen et al. to perform *operando* electrochemistry experiments on various crystal surfaces including Au and Pt[13,14]. Transmission cells have also been used to study gas-solid interfaces under high pressure and moderate temperatures (0-100 bar and -40 to 70ºC), enabling the characterization of $CO_2$ adsorption on polystyrene thin films[15–17].

Ultimately, both the thin-film and the transmission cell design have proven to be advantageous for probing atomic structures and have been integrated to probe a variety of systems including liquid and gas at elevated temperatures and pressures. However, all the reflectivity cell designs discussed above are operable exclusively at synchrotron facilities, where ample hutch space, high incident flux, and beamline expertise enable controlled *in situ* and *operando* studies. Beamlines are often oversubscribed, and even if allocation is successful, waiting times to perform experiments can be on the order of months or longer. In-house lab diffractometers offer a solution to these challenges, and improvements of commercial scattering instruments and diffractometers offer good resolution with quick acquisition time of order tens of minutes[18].

In this work we develop an XRR transmission cell capable of performing temperature-dependent *in situ* vapor dosing experiments on in-house diffractometers. By enabling elevated temperature gas-phase XRR studies at university or industry laboratories, the characterization of complex systems or buried interfaces can be expedited. We developed a vacuum manifold system where vapors of varying volatility are dosed into the cell environment. Ultimately, this allowed for adsorption studies to be performed by collecting reflectivity as a function of chemical potential (gas vapor pressure). We discuss the cell and manifold system design as well as its application to several experiments including the adsorption of water into polyamide (PA) membranes on silicon and temperature-dependent studies of polystyrene (PS) pseudo-brushes on alumina. In this work we focus on reflectivity measurements performed in the range of 0-250mbar and at temperatures ranging from 25 to 200ºC. This methodology may be modified for other grazing incident methodology, such as grazing incidence X-ray diffraction, or gas phase measurements at higher temperatures and pressures.

## II. Cell and Manifold Design

In its simplest form, an air-free reflectivity cell requires air-tight X-ray transparent windows and a method to secure the substrate that allows for grazing-incidence measurements to be performed. A goal for this work was to create a cell capable of holding vacuum (< 0.1 mbar) or vapors of controlled pressure, while simultaneously performing experiments over a range of temperatures. Vacuum capabilities along with a custom manifold system allow for controlled chemical potential studies to be performed through the introduction of precise partial pressures of a vapor. The cell was designed for an in-lab diffractometer but also could be used at synchrotron facilities.

Figure 1 displays the custom cell with a solid view (Figure 1A), a semi-transparent view (Figure 1B), and an exploded view (Figure 1C). The design consists of four main structural and functional components including the stainless-steel cell body (3), the pressure transducer (1), a copper block for heat conduction (9), and the heating cartridge (6). Wires are attached to the heating cartridge but are omitted from the schematics for clarity.

The heating cartridge chosen for these applications has a $^{1}/_{8}$" diameter that allows the cartridge to slide into the cell through the copper block, seen most clearly in Figure 1B. Thus, the cartridge heater acts as a pin which holds the copper block in place when under vacuum conditions. A setscrew (15) is used to hold the cartridge in place. The thermocouple is manufactured inside of the cartridge heater's body, near the far end, and measures the temperature of the copper block.

The high thermal conductivity of copper (401 W/m*K) ensures direct monitoring of the substrate temperature. Direct monitoring of the temperature relies on the assumption that the temperature gradient through a thin (1 mm) sample is small. The cylindrical copper block is inserted with a Viton gasket (11) which ensures the stability of the vacuum seal. Just above this gasket seat, the copper block is cut perpendicular to the face of the copper cylinder, shown in Figure S1. This slit increases the flexibility of the copper block about the empty midsection. An additional depression on the surface of the cylinder has been made to fit a 1cm x 1cm sample and could be varied depending on substrate size and shape. When the sample (12) is placed inside, the setscrew (13) is used to push the copper pieces together – ultimately providing a vise-like grip on the substrate.

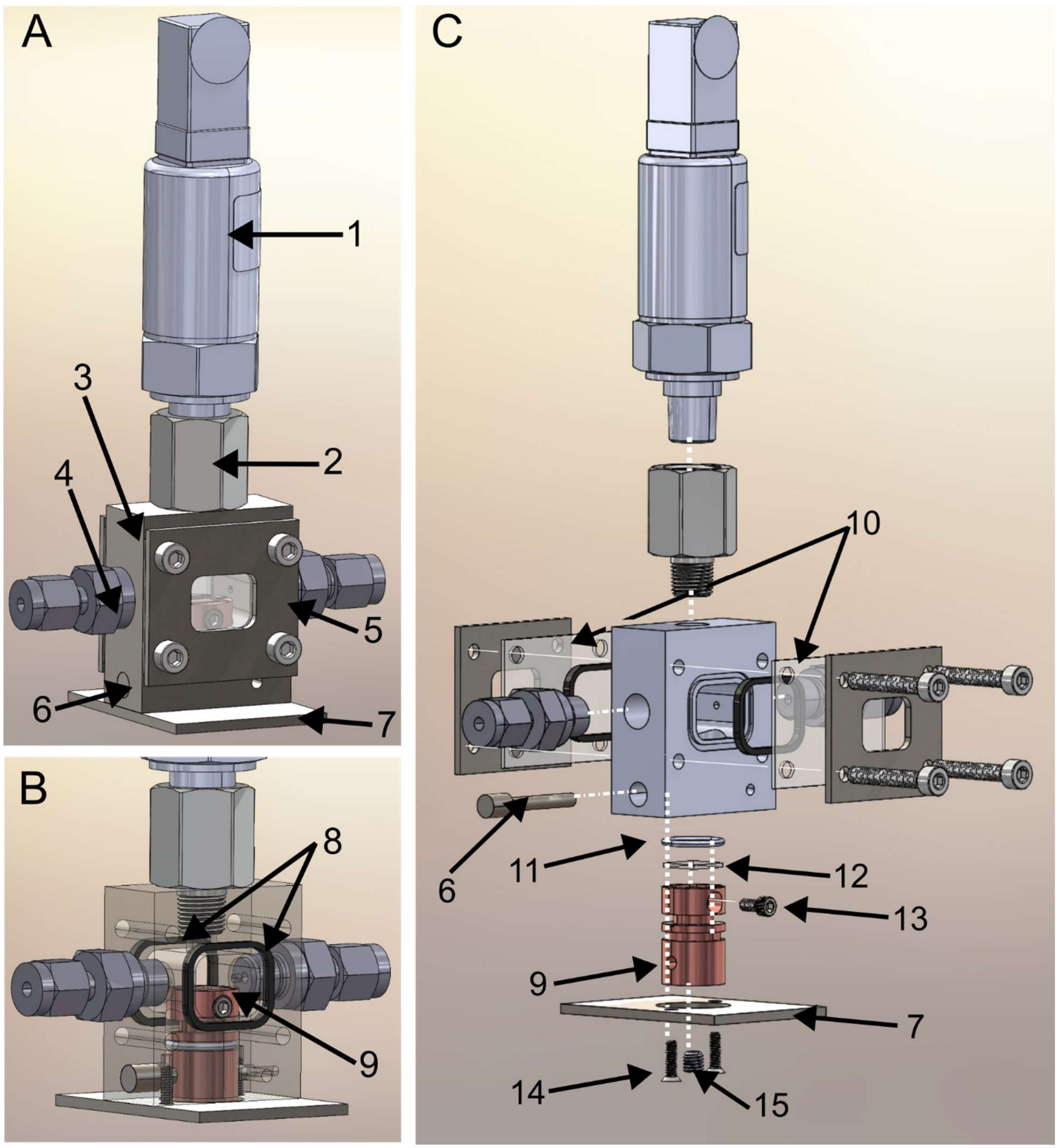

**Figure 1:** *In situ* reflectivity cell design. A) Fully assembled design. B) Fully assembled design with transparent cell body. C) Exploded assembly. 1 – Pressure transducer, 2 – 1/8” to 1/4” NPT adapter, 3 – stainless steel cell body, 4 – 1/8” tubing connections, 5 – stainless steel frame, 6 – heating cartridge, 7 – base plate, 8 – window frame Viton gaskets, 9 – copper block, 10 - PEEK windows, 11 – copper block Viton gasket, 12 – substrate, 13 – sample set screw, 14 – base plate screw, 15 – copper block set screw.

The pressure transducer, which is connected via a $^{1}/_{8}$” to $^{1}/_{4}$” NPT adapter (2), allows for direct monitoring and logging of the internal cell pressure with a precision of +/- 1 mbar. The USB cable, which extends from the transducer, has been omitted in Figure 1. Two side ports for $1/8$ " tubing connections are located near the center of the side body (4) and allow for attachment to the vacuum manifold system. The second port allows liquid or gas flow-through experiments. Optionally, one port can be capped if fluid flow past the sample is not needed. A detailed drawing of the cell and copper stage is illustrated in Figures S2 and S3. To ensure proper sealing of the sample, we utilize Viton gaskets on the external body of the cell (8) for the attachment of thin (0.7 mm) polyether ether ketone (PEEK) windows (10), which are held in place by threaded stainless steel frames (5). The PEEK windows reduce transmission by ~65%, which may be improved by using Kapton windows with 50μm thickness (Figure S4). However, the PEEK windows were found to significantly improve vacuum seal capabilities with signal-to-noise that is acceptable up to $Q_z \sim 0.6$ Å$^{-1}$. The cell is then mounted to a block which directly attaches to the diffractometer's stage (7), making removal and replacement of the cell simple.

A requirement when designing the cell and manifold attachment system for the in-lab diffractometer was ease of removal and replacement. Since many users require the diffractometer, and the space is rather small, it is critical that limited bulky hardware be permanently attached inside the X-ray enclosure. Additionally, any hardware that remains inside of the diffractometer must remain far from source and detector arms to avoid crashes during translation.

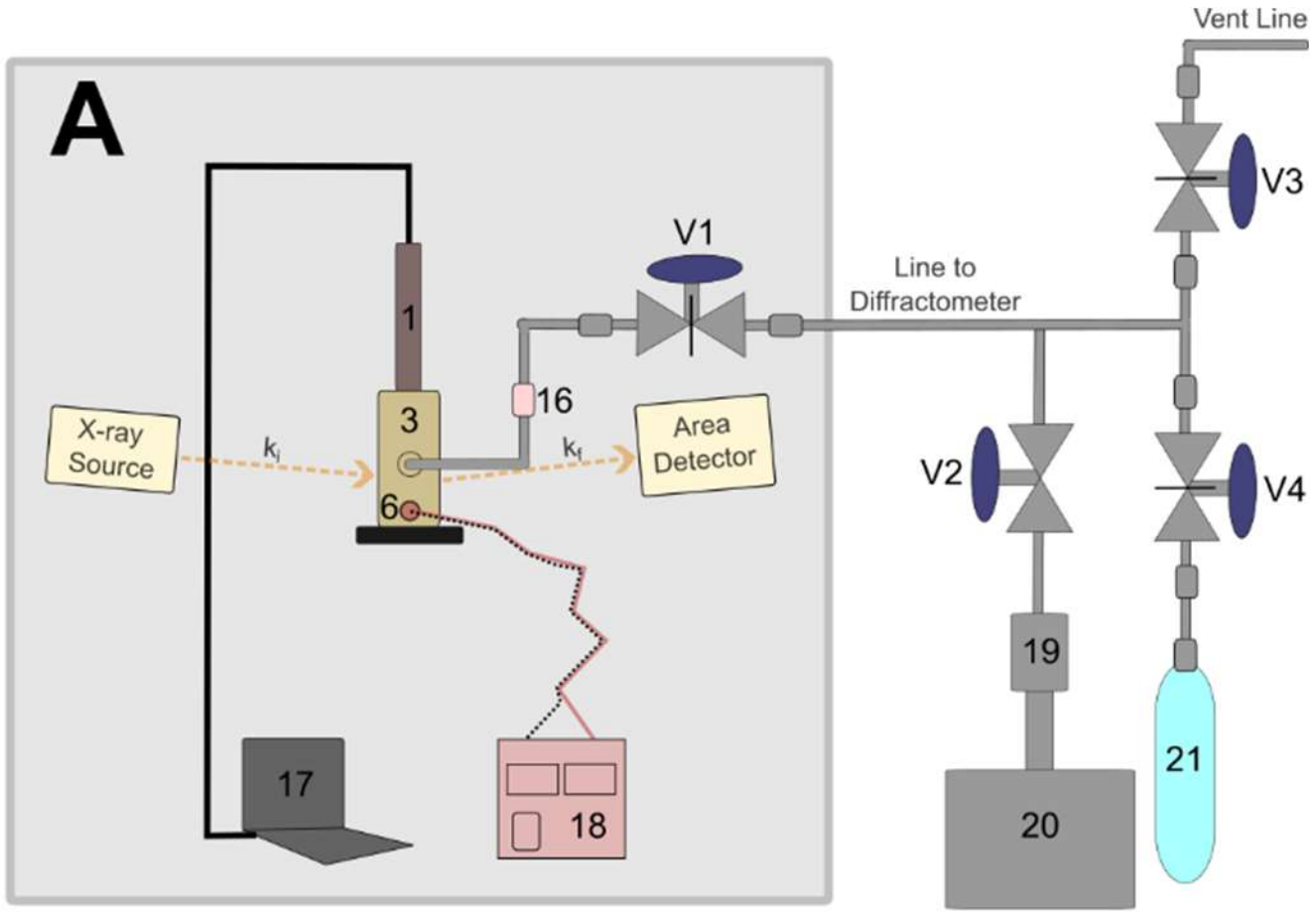

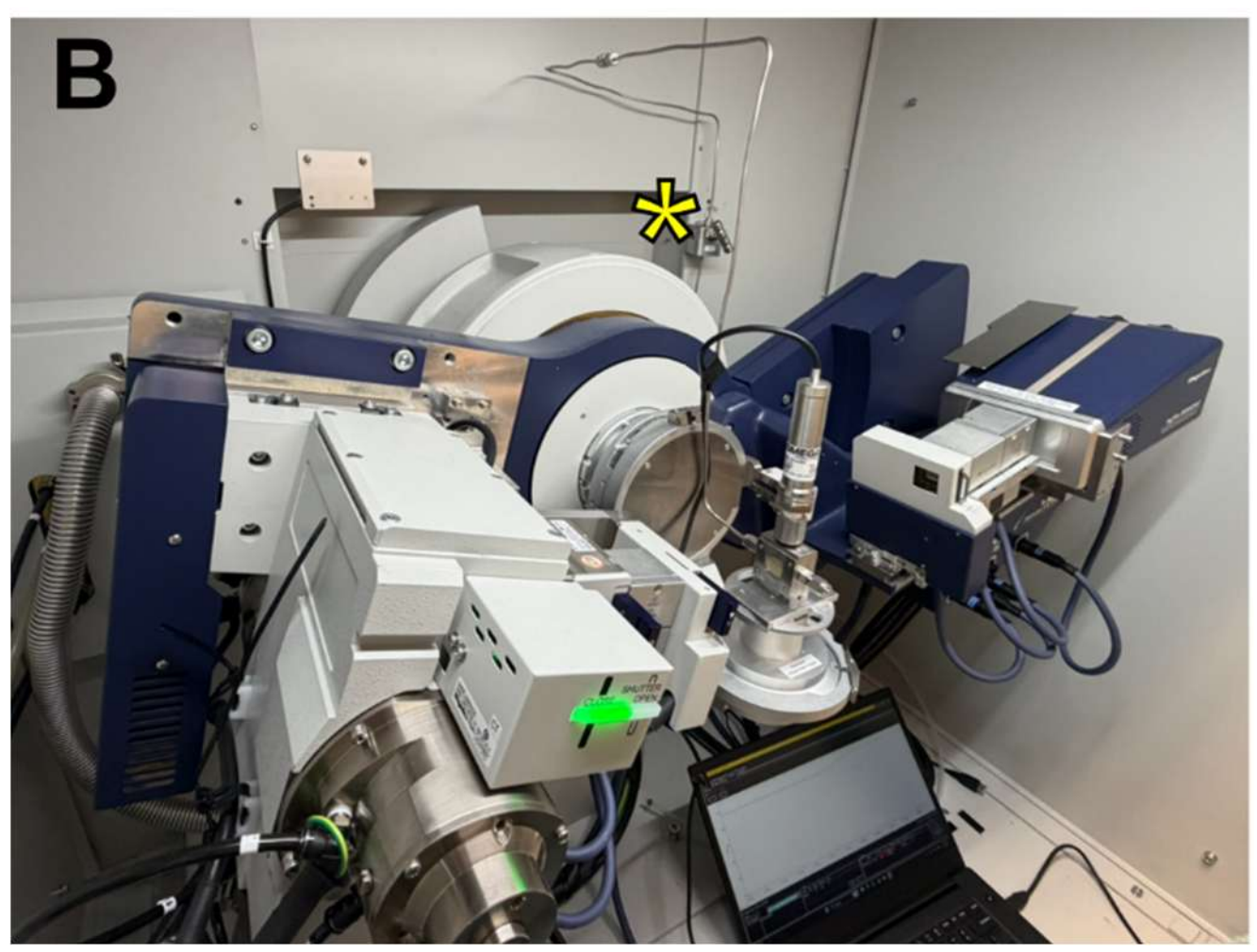

**Figure 2:** Manifold system for vapor dosing. A) Schematic of manifold system displaying the system inside of the diffractometer (gray shaded box) and outside of the diffractometer. (1) Pressure gauge, (3) cell body, (6) heater, (16) critical junction for cell removal, (17) computer, (18) temperature controller, (19) oil trap, (20) vacuum pump, (21) glass beaker. The X-ray source and area detector are shown explicitly with momentum transfer vectors ($k_i$ and $k_f$) to clarify reflection geometry. Wires extending out of the heater (6) signify the power cord (red line) and thermocouple (black line). V1, V2, and V3 are needle valves, designated with a line bisecting the valve. V2 is a gate valve. B) Photo of system set up inside of the diffractometer. A yellow asterisk highlights the throttle valve (V1).

Thus, much of the manifold system is located outside of the diffractometer, as shown by the process flow diagram in Figure 2A. The vacuum pump (20) and the probe compound (21) are connected to the main manifold structure. In the context of this work, the probe compound is a liquid, contained inside of a glass beaker. A lecture bottle or gas cylinder may be exchanged with the glass beaker to enable experimentation with gases at elevated pressures.

For this setup, a total of four valves were required to control vacuum flow (V2), bleed to air (V3), introduce probe compounds into the system (V4), and tune the partial pressure inside of the cell (V1). Only the throttle valve (V1) sits inside the diffractometer enclosure and is positioned out of mechanical interface range of any arm movement. A magnet was used to attach the valve to the diffractometer wall, eliminating the need to bore holes in the X-ray shielding. This valve is highlighted with an asterisk in Figure 2B.

The tubing which extends from V1 travels towards the cell and close to the enclosure wall is positioned to avoid interrupting arm translation. The tubing reaches the middle of the diffractometer before extending towards the cell. To avoid reinserting the tubing directly connected to V1 each time a user performs measurements, a $^1/_8$" x $^1/_8$" adapter (16) was placed between the sample stage and the manifold. This streamlined feature improved the stability of the vacuum system as critical seal deformation was minimized (i.e. at the V1 connection).

The temperature controller and computer used to connect the pressure transducer also sit inside of the diffractometer during experimentation. It should be noted that having the computer inside of the diffractometer allows for direct inspection of the cell pressure when throttling vapor into the cell.

When utilizing a liquid probe, the headspace and any dissolved gas must be removed from the glass bottle (21). This is done by first evacuating the entire manifold systems while V3 and V4 remain closed. Shielding the roughing pump via V2 allows for opening of V4 and exposure of the glass bottle to vacuum pressures. Finally, V4 is closed again before the system is vented to atmosphere via V3. This cycle of vacuum pumping → liquid exposure → purging is repeated until all headspace and dissolved gas has been removed from the bottle, when the saturation pressure of liquid is reached. We have found that this requires anywhere from 10-15 cycles (Figure S5).

This procedure is performed prior to reflectivity collection. It is critical to note that if a sample has been pre-loaded into the cell before liquid cycling, the throttle valve (V1) should remain closed as to not prematurely expose vapor to the sample. After proper probe molecule isolation and installation of a sample inside of the cell, reflectivity measurements begin.

A standard example of a vapor dosing experiment is now discussed. The manifold and cell systems are evacuated such that valves V1 and V2 are open and valves V3 and V4 are closed. Sample alignment takes place followed by collection of XRR. We denote this measurement as 0 mbar, as the cell is fully exposed to the base pressure of the roughing pump (< 0.1 mbar). Following data collection V1 and V2 are closed. V3 is opened, thus exposing the manifold system outside of the diffractometer to the saturation pressure of the liquid. If it is desirable to perform stepwise pressure studies, the throttle valve inside of the diffractometer is then briefly cracked open and immediately closed. Another alignment and reflectivity scan can then be collected. For stepwise measurement this procedure is repeated until the saturation pressure is reached, or when the V1 throttle valve can remain open without pressure increase. Reversibility studies can also be performed by closing V1 and V3 and opening V2 to bring the manifold system back to base pressure. Throttling V1 now results in a controllable pressure decrease inside of the cell system.

## III. Experimental Results

### A. Vapor Dosing Studies

To showcase the versatility of the cell system we tested volatile probe molecules including heptane, 2,3-dimethylbutane (DMB), and methanol (Figure 3A) with no sample inside of the cell. Performing standard vapor dosing measurements in this way allows for stability to be more precisely understood – since there is no film readily absorbing vapor and thus changing cell pressure. It should be clarified that the pressure measurements in Figure 3 do not correspond to specific XRR measurements and were performed specifically to evaluate the stability and equilibration time for various probe molecules.

Vapor was throttled into the cell incrementally via opening and immediate closing of V1 and then allowed to equilibrate for ~1 minute before throttling to the next pressure. The target pressure for each step was predefined as to allow for between 8-10 pressure increments between base vacuum pressure and the maximum vapor pressure of each compound. After the maximum vapor pressure was achieved, pump down experiments were performed. The average vapor pressure is calculated every minute and expressed as a relative percent, compared to the saturation vapor pressure (Figure 3B,C, and D) for each probe molecule. Resulting error bars correspond to the standard deviations of the average values obtained per minute. About 10 pressure points are recorded per second (~10 Hz), thus every scattered point and error is reflective of ~600 data points. Generally, we find good equilibration of the cell on the one-minute time scale. However, for more volatile liquids (2,3-DMB), larger errors persist and indicate that longer equilibration time is required. The increased error is likely a manifestation of the pressure transients experienced when opening the throttle valve (V1), which are larger for the more volatile species. Averaging over the course of each minute includes these pressure transients and thus results in a higher error. In addition to showcasing the use of various probe molecules, these results also highlight the ability of the manifold system to achieve the saturation pressure. For all cases we can achieve 90-95% saturation conditions.

Reliable pressures and a streamlined approach make the cell adaptable and effective for performing pressure experiments with liquid probe molecules of varying volatility using an in-lab X-ray source. The manual control of V1 is a limitation of the current cell design and may be improved by replacement with a stepper motor valve. Introduction of a stepper motor would enable more precise and stable pressure control. We also speculate that a controlled motor would also allow for highly saturated (98-99%) environments to be consistently achievable.

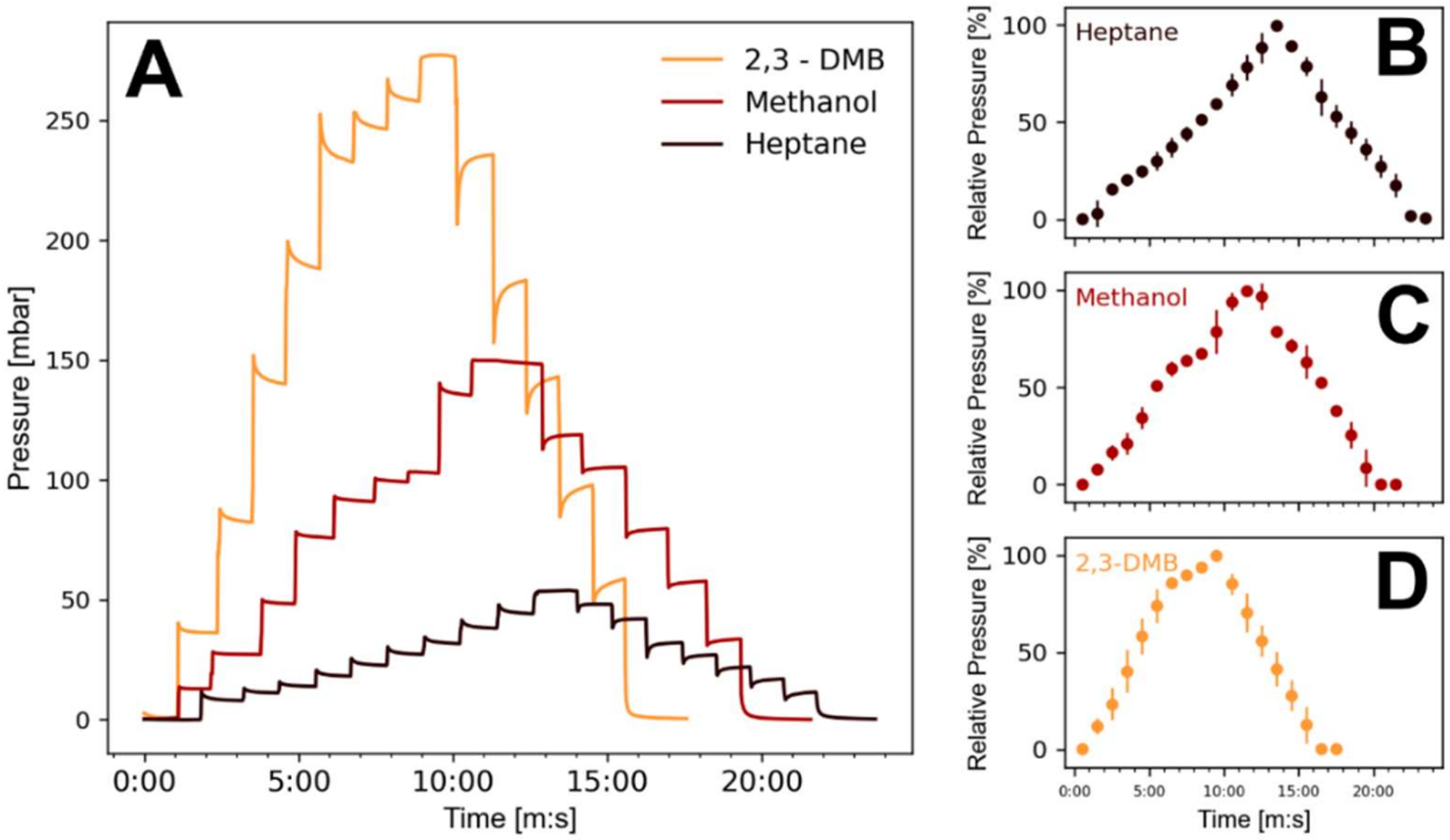

**Figure 5:** A) Pressure ramping profiles for 2,3 – DMB (orange), methanol (red), and heptane (black) as a function of time. Relative pressure of B) heptane, C) methanol, and D) 2,3-DMB as a function of time. Scatter points are taken by averaging the recorded pressure every minute with standard deviations represented by error bars.

To showcase the ability of this dosing manifold system for use while performing XRR studies, we characterize the swelling of polyamide (PA) membrane films on silicon. Distinct oscillations, known as Kiessig fringes, present at 0 mbar are characteristic of a film with definitive thickness and electron-density (Figure 4A). Upon vapor exposure, the minimum of Kiessig fringes shift towards smaller $Q_z$, indicative of a thickness increase associated with membrane swelling. Inspection of the membrane thickness from the oscillation frequency prior to water vapor adsorption experiments (0 mbar) and at maximum vapor pressure (30 mbar) further elucidates the impact of water vapor on the PA film (Figure 4C). Dividing 2π by the difference in the first two minimum positions ($\Delta Q_z$) allows for an estimation of thickness ($d = 2\pi/\Delta Q_z$). Estimated thickness values are extracted from each reflectivity curve in this way (Figure 4B), showing the total film swelling is approximately 1.6 nm. A smoothing spline was fit to the data to guide the eye and highlight the overall thickness changes as a function of temperature. Reflectivity can also be more rigorously fit using the Parratt formalism to extract more accurate thickness, roughness, and electron density changes[19]. We have found that this system is effective for identifying changes on the order of < 0.5 Å.

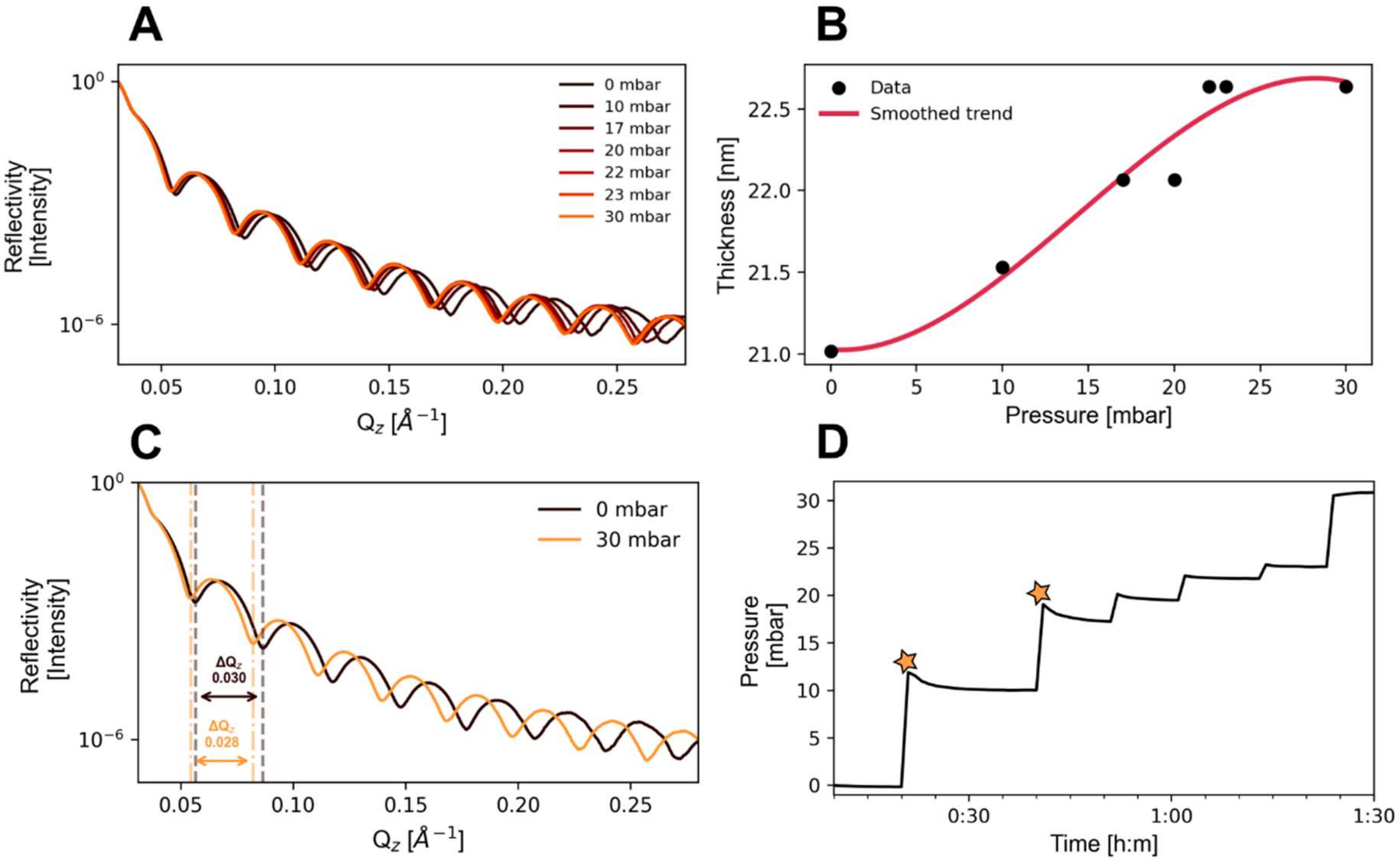

**Figure 4:** A) *In situ* reflectivity of PA membranes in contact with water vapor at various vapor pressures. Colors correspond to the respective pressure ranging from 0 to 30 mbar. B) Extracted thickness in nm vs. pressure of PA membranes. A smoothing spline was fit the data to guide the eye and highlight overall trends and is shown with a red line. C) XRR of PA membranes under rough vacuum conditions (0 mbar – black) and at maximum vapor pressure (30 mbar – orange). D) Cell pressure versus time for the duration of sample measurements. Orange stars denote pressure transients upon throttling.

Examining the pressure profile obtained during these experiments (Figure 4D), allows for confirmation of pressure stability. We report pressure transients when vapor is initially throttled, which results in a decay to steady state. These transients are denoted with orange stars in Figure 4D. Each reflectivity scan was performed within 10 minutes. Repeat experiments at 10 mbar were collected to ensure the membrane was reaching an equilibrated state within this time scale (Figure S6).

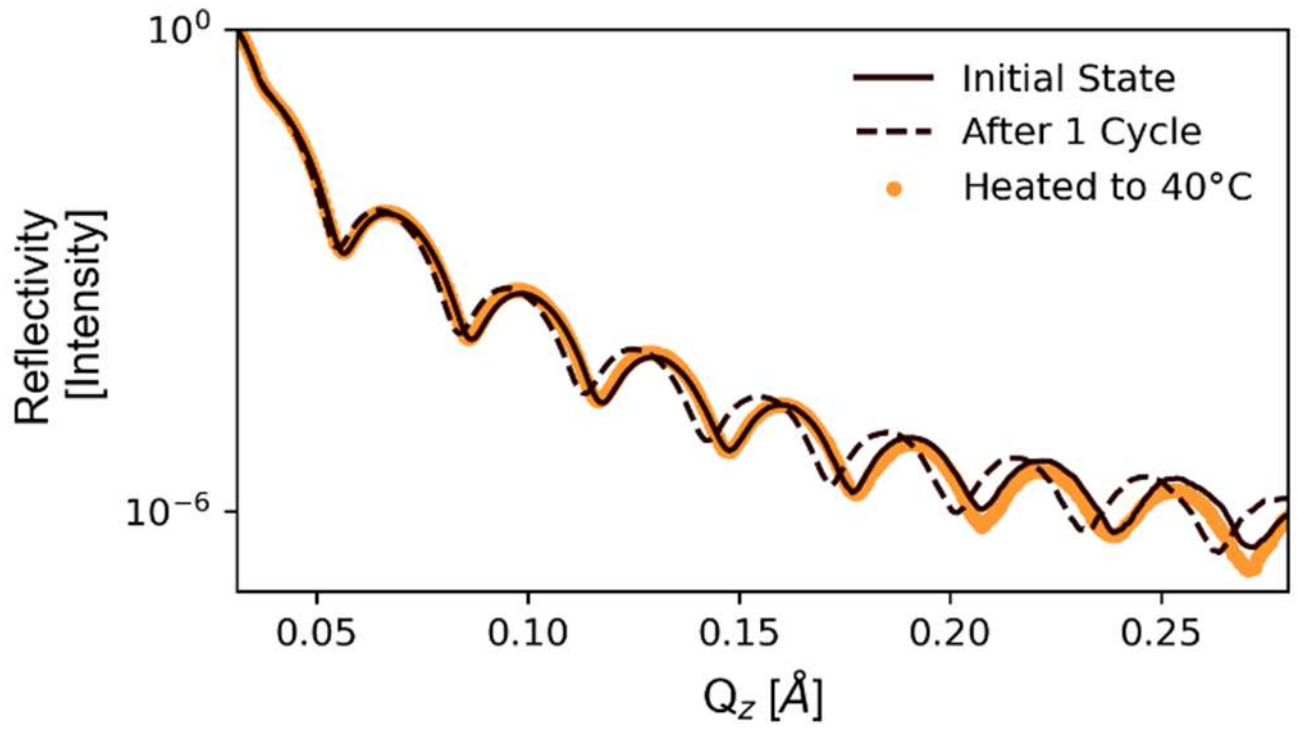

**Figure 5:** Reflectivity obtained for PA films at 0 mbar prior to vapor exposure (solid lines), after exposure to maximum vapor pressure (dashed lines) and after the sample was heated to 40ºC (orange dots).

Following pressurization to near-saturation conditions, the cell is brought back to rough vacuum, constituting one "cycle" of vapor exposure. The reflectivity profile after one cycle is shown in Figure 5. The minimum of Kiessig fringe oscillations remains shifted compared to its initial state, indicating that some water vapor cannot readily desorb from the membrane on the experimental timescale. However, when the cell is heated to 40ºC and held at rough vacuum, shown as orange dots in Figure 5, the PA sample effectively returns to its initial state. This confirms that some water is bound more strongly within in the membrane and requires additional thermal energy to escape on the time scales of the experiment.

### B. Temperature Dependent Studies

As described previously, this cell also enables X-ray measurements during *in-situ* heating. A cartridge heater directly heats the copper block inside the cell (Figure S7). Although the heater itself reaches 1260 °C, the PEEK windows soften near 343 °C. Thus, measurements are kept well below this limit. To demonstrate the combined capability, we measured polystyrene (PS) pseudo-brushes on single-crystal alumina while heating (Figure 6a) and dosing with a toluene (Figure 6b). Here, the toluene has strong affinity for the polymer, impacting the polymer structure resolved in the measurement. Figure 6 shows the Fresnel-normalized reflectivity, $R/R_F$, where $R_F$ is the theoretical reflectivity of an ideally flat and sharp interface. Normalizing this way suppresses the steep decay in $R$ and makes the fringes easier to visualize.

For the temperature ramp, the cell was held under rough vacuum (< 0.1 mbar) to prevent adsorption of contaminants. At each setpoint the sample equilibrated for 15 minutes before measurement. Sample alignment was performed following each ramp to maintain proper reflected signal following thermal expansion of the cell.

From these measurements, we tracked how the pseudo-brush thickness evolved across the temperature ramp. Brush thicknesses were estimated from the first Kiessig fringe minimum ($d \sim \pi/Q_{min}$). Because the pseudo-brushes are so thin, the in-house diffractometer resolves only this first fringe at elevated temperatures, so we chose to use only the first minimum to determine thickness. At 25 °C the reflectivity shows well-defined fringes, resulting in an initial pseudo-brush thickness of ~40.7 Å. When heated to 100 °C, the fringe shifts towards higher $Q_z$, corresponding to a decrease in thickness to ~35.7 Å. Films this thin undergo their glass transition below the bulk value of 100 °C, generally around 75 °C[20,21]. Heating past this temperature allows for the brush to transition from a glassy, rigid state into a rubbery state with mobile chains, allowing the chains to reorganize and resulting in a thinning of the entire pseudo-brush. Heating to 200 °C results in dampening of the fringe, and the signal becomes dominated by surface roughness rather than a well-defined film thickness.

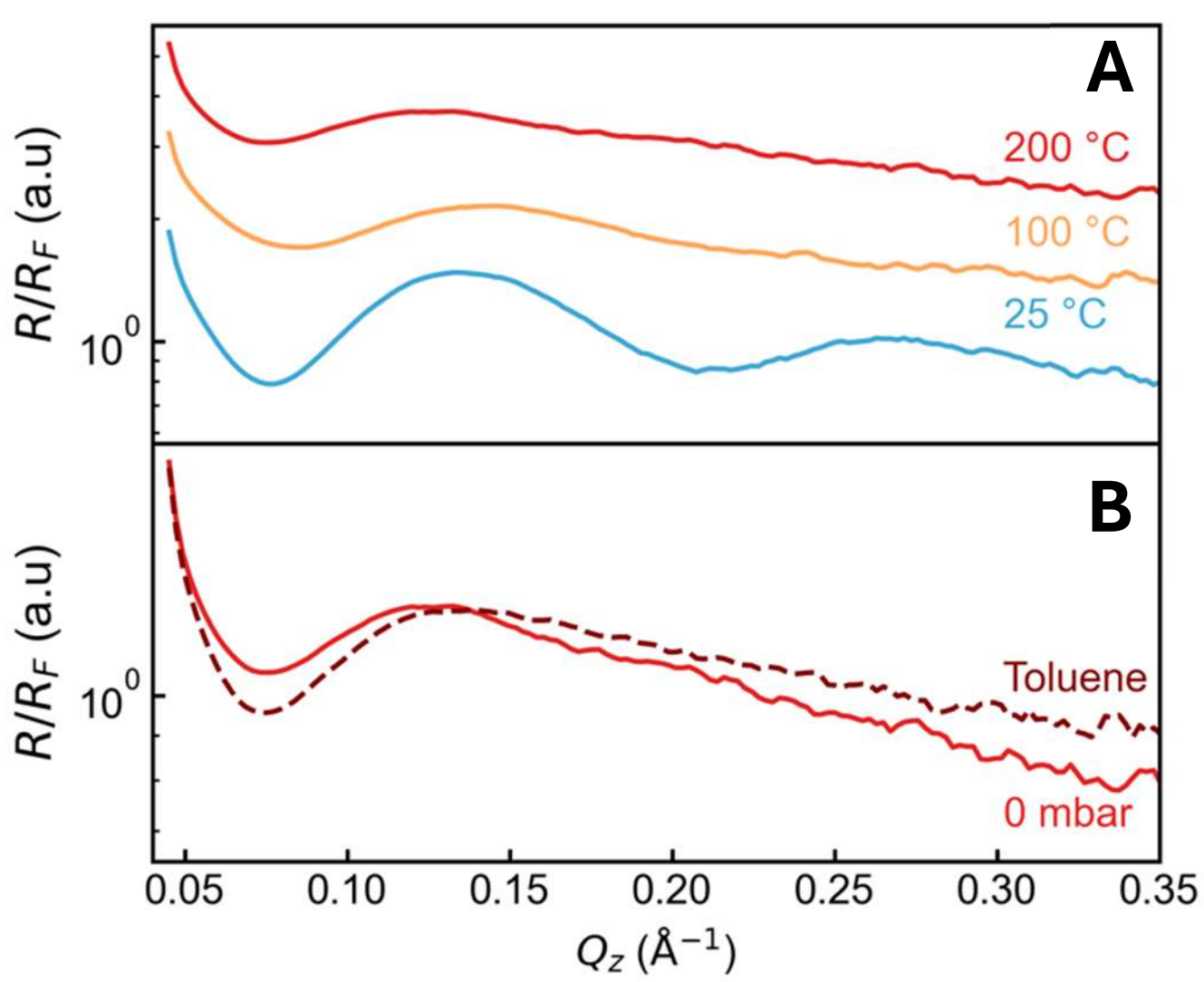

**Figure 6:** Fresnel-normalized reflectivity collected for PS pseudo-brushes. A) In vacuum at 25 ℃ (blue), 100 ℃ (yellow), and 200 ℃ (red). B) At 200 ℃ in vacuum (solid line) and toluene vapor dosing (dashed line).

Introducing saturated toluene vapor (38 mbar) at 200 °C produced no measurable change in pseudo-brush thickness, but the amplitude of the reflectivity oscillations increased, indicating greater electron density contrast. We attribute this to favorable intercalation between toluene into the PS brush. The toluene fills the free volume and raises the electron density, enhancing the reflectivity contrast while preserving the overall thickness. We found that the dosed state of the PS was distinct from the initial dry brush at room temperature (Figure S8).

The heating experiments discussed demonstrate the capabilities of the cell: resolving Ångstrom scale structural changes in a surface adsorbed polymer under both thermal and chemical stimuli in a single *in-situ* X-ray experiment. However, there is currently no way to directly control the temperature of probe molecules before it enters the cell. Heat tracing the glass beaker and manifold systems would provide an effective method for ensuring a constant temperature across all systems, enabling better control of saturation pressure by changing probe molecule temperature and eliminating any cold spots in the cell.

## IV. Conclusion

An *in situ* reflectivity cell and manifold system to enable temperature-controlled vapor dosing experiments using in-house lab diffractometers is presented. The streamlined design allows for ease of removal and attachment, ideal for a system with many users. Additionally, solvent exchange requires little to no effort, allowing higher throughput experimentation. The transmission reflectivity cell designed for these experiments offers simple yet robust capabilities, enabling vacuum-tight operation and temperature studies. We highlight the vapor dosing capabilities via water adsorption studies into polyamide films and polystyrene pseudo-brushes heated to 200ºC to exemplify temperature capabilities.

## V. Materials and Measurements

The digital pressure transducer utilized for these experiments was purchased from Omega (PX409-015AUSBH) and has an accuracy of ±1 mbar. Cartridge heater with embedded thermocouple was purchased from McMaster Carr (part number #8440T25). Vacuum pressure pump is an AMEB 71 NX2 type pump produced by AEG.

PA films were synthesized using a custom-built automated system (MDUSA, molecular deposition using spin assistance) for controlled solution-based molecular layer-by-layer deposition at the University of Houston. The PA is a polymer formed from 40 MDUSA cycles reacting m-phenylenediamine with trimesoyl chloride monomers. The polymer was adhered to silicon wafers using the functional amino-silane agent 3-aminopropyl)triethoxysilane. The method for creating these films is detailed elsewhere[22]. PS pseudo-brushes were deposited onto single-crystal alumina by spin-coating. To isolate the pseudo-brush, the samples were washed in toluene for an extended period of time. This procedure follows previously reported methods[6,7].

Reflectivity was performed using a Rigaku Smartlab X-ray diffractometer with a Cu K-α source. Standard 0D reflectivity was used, and detector integration was performed using Smartlab software. The reflectivity was recorded from 0 to 0.7 $Å^{-1}$ with a scan rate of 0.14 $Å^{-1}$ / minute.

**Acknowledgements**

This work was supported and funded by the National Science Foundation Office of Emerging Frontiers in Research and Innovation (NSF-2132033). Data was collected using the Rigaku Smartlab X-ray Diffractometer, an instrument operated and maintained by COSINC facilities. We additionally thank Dr. Nicholas Weadock for his support with instrument operations. The cell design was executed by Dragan Mejic, and through his machining expertise these measurements were possible.

**Author Contributions**

*E.E.D.*: conceptualization; formal analysis; investigation; methodology; writing – original draft; writing – review & editing. *M. A. F.:* investigation; methodology; sample preparation; writing – review & editing. *S.R.N.*: conceptualization; investigation; sample preparation; methodology; writing – review & editing. *J. M. W.:* conceptualization; methodology, writing – review & editing, funding acquisition. *M. F. T.:* conceptualization; methodology, writing – review & editing.